\documentstyle[psfig,conf_iap]{article}
\newcommand{\apjs}{ApJS\,\,}
\newcommand{\lya}{Lyman-$\alpha$\ }
\newcommand{\la}{\ \raise -2.truept\hbox{\rlap{\hbox{$\sim$}}\raise5.truept
        \hbox{$<$}\ }}
\newcommand{\ga}{\ \raise -2.truept\hbox{\rlap{\hbox{$\sim$}}\raise5.truept
        \hbox{$>$}\ }}

\newcommand{\be}{\begin{equation}}
\newcommand{\ee}{\end{equation}}
\newcommand{\ba}{\begin{eqnarray}}
\newcommand{\ea}{\end{eqnarray}}
\begin{document}
\heading{%
%
Clustering of Absorbers
%
} 
\par\medskip\noindent
\author{%
S. Cristiani$^{1}$, S. D'Odorico$^{2}$, V. D'Odorico$^{3}$, 
A. Fontana$^{4}$, E. Giallongo$^{4}$, L. Moscardini$^{1}$, S. 
Savaglio$^{2}$
}
\address{%
Dipartimento di Astronomia dell'Unversit\`a di Padova, Vicolo
dell'Osservatorio 5, I-35122 Padova, Italy
}
\address{%
European Southern Observatory, K.Schwarzschild-Strasse 2, D-85748
Garching bei M\"unchen, Germany } \address{%
International School for Advanced Studies, SISSA, via Beirut 2-4,
I-34014 Trieste, Italy } 
\address{%
Osservatorio Astronomico di Roma, via dell'Osservatorio, I-00040 Monte
Porzio, Italy }

\begin{abstract}
The observed clustering of Lyman-$\alpha$ lines is reviewed and
compared with the clustering of CIV systems. 
We argue that a continuity of properties
exists between Lyman-$\alpha$ and metal systems and show that the
small-scale clustering of the absorbers is consistent with a scenario
of gravitationally induced correlations. At large scales 
statistically significant over and
under-densities (including voids) are found on scales of tens of Mpc.
\end{abstract}
\section{Introduction}
 
The search for clustering of the Lyman-$\alpha$ lines has produced
along the years diverse results.  Systematic studies of the
distribution of redshifts in the QSO Lyman-$\alpha$ forest began in
the early 80's with the work by Sargent et al.  \cite{sarg80} who
concluded that no structures could be identified.  Almost all the
subsequent results have failed to detect any significant correlation
on velocity scales $300 < \Delta v < 30000$ km s$^{-1}$
\cite{sarg82,becht87,webb91}.  On smaller scales ($\Delta v =50-300$
km s$^{-1}$) there have been indications of weak clustering
\cite{webb87, rauch92, cherno95, cris95, meiks95}, together with
relevant non-detections \cite{pet90, eric93, Kirk97}.

On the contrary, metal-line systems selected by means of the CIV
doublet have been early recognized to show strong clustering up to 600
km s$^{-1}$ \cite{sarg88}, suggesting a different spatial
distribution.  In fact, the absence of power in the two-point
correlation function has been claimed as a striking characteristic of
the Lyman-$\alpha$ forest and has been used as a basic argument to
develop a scenario of the Lyman-$\alpha$ absorbers as a totally
distinct population with respect to metal systems and therefore
galaxies.

\section{The Lyman-$\alpha$ database}
Our analysis of the clustering of the Lyman-$\alpha$ lines is based on
data obtained in the framework of an ESO key-programme devoted to the
study of QSO absorption systems at high redshifts.  We have obtained
spectra of several QSOs, with emission redshifts ranging from 3.27 to
4.12, at resolution between 9 and 14 km s$^{-1}$ \cite{cris97}.  We
have complemented our data with other spectra available in the
literature with similar resolution and redshift range,
\cite{carsw91,rauch93,1700,Kirk97,Hu95} obtaining a final sample of 14
QSOs (in the following {\it extended sample}), more than 1600 \lya's,
a suitable database to investigate clustering, especially at high
redshift, where the density of lines provides particular sensitivity.

\section{Small-scale clustering of Lyman-$\alpha$ lines}
To study the clustering properties of our sample of Lyman-$\alpha$
lines we have adopted the two-point correlation function (TPCF) in the
velocity space:
\begin{equation}
\xi(\Delta v) = {N_{obs}(\Delta v) \over N_{exp}(\Delta v)} -1
\end{equation}
In our case sample $N_{exp}$, the number of pairs expected in a given
velocity bin from a random distribution in redshift, is obtained
averaging 1000 numerical simulations of the observed number of
redshifts, trying to account for all the relevant cosmological and
observational effects.  In particular the set of redshifts is randomly
generated in the same redshift interval as the data according to the
cosmological distribution $\propto (1+z)^{\gamma}$, where the best
value of $\gamma = 2.65$ has been derived from a maximum likelihood
analysis \cite{gial96}.  The results are not sensitive to the value of
$\gamma$ adopted.  Incomplete wavelength coverage due to gaps in the
spectrum or line blanketing of weak lines due to strong complexes is
also accounted for.  Lines with too small velocity splittings,
compared with the finite resolution or the intrinsic blending are
excluded in the estimate of $N_{exp}$.  The resulting correlation
function for the full {\it extended sample} of Lyman$-\alpha$ lines
shows a weak but significant signal, with $\xi\simeq 0.2$ in the 100
km s$^{-1}$ bin: 739 pairs are observed while only 624 are expected
for a random distribution, a $4.6 \sigma$ deviation from
poissonianity.

\begin{figure}[t]
\centerline{\vbox{ \psfig{figure=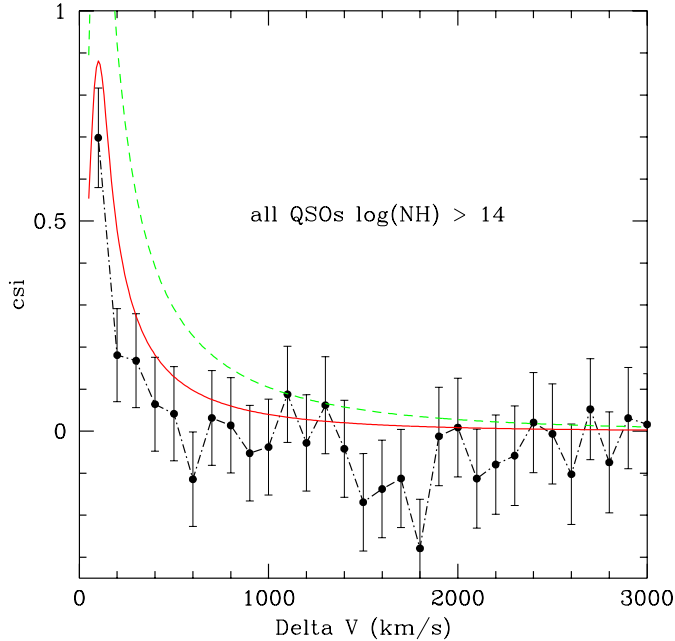,height=9.cm} }}
\caption[]{\label{fig1} Two-point correlation function in the velocity
space for lines with column densities $> 10^{14}$ cm$^{-2}$.  Errors
estimated with bootstrap resampling.  The continuous line corresponds
to a CDM model normalized to reproduce the observed abundance of
clusters, the dashed line corresponds to a tilted CDM (see section 5).
}
\end{figure}

We have then explored the variations of the clustering as a function
of the column density.  For lines with $\log N_{HI}\le 13.6$ no
evidence for clustering is present.  On the contrary, for lines with
$\log N_{HI}\ga 14$ (Fig. \ref{fig1}), the correlation function at
$\Delta v = 100$ km s$^{-1}$ shows a remarkable increase in amplitude
($\xi\simeq 0.7$) and significance: 234 pairs are observed while only
145 are expected for a random distribution, a more than $7 \sigma$
deviation from poissonianity.  No relevant feature other than the peak
at small velocity separations is observed. In particular previous
claims for anti-clustering on scales $\sim 600 - 1000$ km s$^{-1}$ are
not confirmed.  Fig.~\ref{figclustew} shows more in detail the
variation of the amplitude of the two-point correlation as a function
of the column density.  A trend of increasing amplitude with
increasing column density is apparent.

We have also studied the evolution of the TPCF with the redshift for
the sub-sample of Lyman-$\alpha$ lines with column densities
$\log(N_{HI}) > 13.8$.  The amplitude of the correlation at $100$ km
s$^{-1}$ decreases with increasing redshift from $0.85\pm0.14$ at
$1.7<z<3.1$, to $0.74\pm0.14$ at $3.1<z<3.7$ and $0.21\pm0.14$ at
$3.7<z<4.0$.  Unfortunately, HST data are still at too low-resolution
or are too scanty \cite{HSTKP7,3C273} to allow a meaningful comparison
with the present data.  Nonetheless, the result by Ulmer \cite{ulmer},
who measured with FOS data a $\xi = 1.8^{+1.6}_{-1.2}$, on scales of
$200-500$ km/s at $0 < z < 1.3$, suggests that the trend persists at
lower redshifts.

\section{Comparison with CIV clustering and discussion}

\begin{figure}
\centerline{\vbox{ \psfig{figure=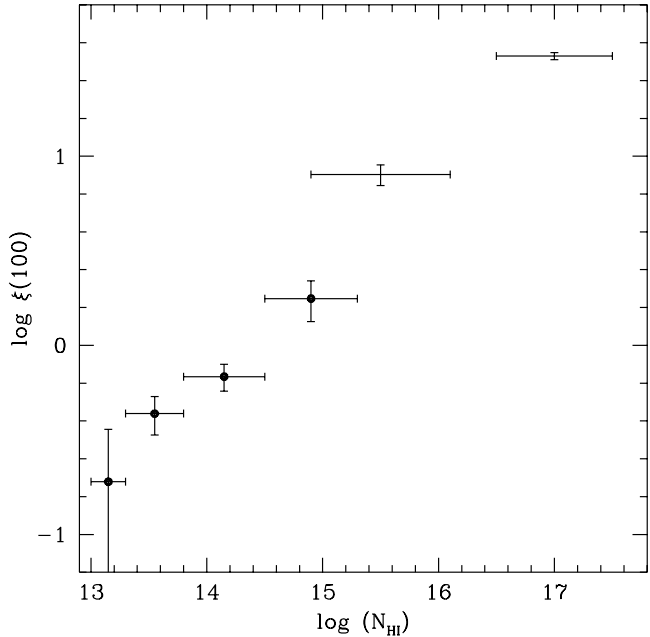,height=9.cm} }}
\caption[]{ Variation of the amplitude of the two-point correlation
function as a function of the column density threshold for the sample
of the Lyman-$\alpha$ lines (filled circles).  The two points in the
upper-right side of the picture show the correlation of the CIV metal
systems (see text)} \label{figclustew}
\end{figure}

A discussion of the clustering of the CIV systems can be found in the
paper by V.D'Odorico (this conference).  A correlation of the
clustering amplitude with column density is observed also for CIV
systems. 
In the upper-right of Fig.~\ref{figclustew} the TPCF
derived for high and low column density CIV metal systems
is shown.  An
extrapolation of the increasing amplitude trend observed for the TPCF
of the Lyman-$\alpha$ lines would easily intercept the corresponding
estimates derived from the CIV metal systems. The similarity between
the shapes of the TPCF's of the Lyman-$\alpha$ and CIV systems, when
observed at comparable resolution \cite{CIVpetit,son:cow:96}, is also
remarkable.

Fernandez-Soto et al. \cite{FS96} have investigated the clustering
properties of the Lyman-$\alpha$ clouds on the basis of the
corresponding CIV absorptions, suggesting that CIV may resolve better
the small-scale velocity structure that cannot be fully traced by
Lyman-$\alpha$ lines. As a consequence, the estimates of the TPCF of
the Lyman-$\alpha$ absorbers, although the effects of the
non-negligible width of the lines (the ``line-blanketing'') are
considered in our simulations, should be regarded as a lower limit to
the real clustering amplitude.  However, it is not straightforward to
translate the properties of the CIV absorbers in the corresponding
ones of the Lyman-$\alpha$ absorbers, since observations show that the
velocity structures of high and low-ionization species are often
different. In any case,  
since the underestimation of the Lyman-$\alpha$ TPCF would
be more severe at larger column densities, the trend of an increasing
correlation with increasing column density appears to be real, from
the lowest column densities up to those corresponding to the strongest
metal systems.

If we add to this observation the following pieces of evidence:
\begin{enumerate}
\item At low redshift a considerable fraction of the Lyman-$\alpha$
lines has been observed to be associated with luminous galaxies and
the local large-scale structure \cite{Lanz95,lebrun96};
\item Metallicities of the order $10^{-2}$, i.e.  similar to the ones
derived for the heavy elements absorptions originated in galactic
halos, are observed for Lyman-$\alpha$ clouds with $\log N_{HI} > 14$
\cite{cowie95};

\item The volume density and cosmological evolution of the same $\log
N_{HI} > 14$ clouds are similar to those of the damped systems
\cite{gial96};
\end{enumerate}
all together this suggests a physical association between the
Lyman-$\alpha$ clouds with $\log N_{HI} > 14$ and the halos of
protogalactic systems.

The typical column density below which no clustering of the
Lyman-$\alpha$ lines is observed ($\log (N_{HI}) \sim 13.6$)
corresponds to the position of the break in their column density
distribution \cite{petit93,gial96}, which has been identified with the
transition from a variety of systems in various stages of
gravitational infall and collapse (or even under-densities) to gas
associated with star forming galaxies \cite{anninos96,mucket96}.

The observed clustering properties are qualitatively consistent with a
scenario of gravitationally induced correlations.  The dependence of
the clustering amplitude on the column density arises naturally in
models involving biasing for the formation of structures in the
universe: objects associated with the stronger potential wells are
expected to be more clustered and the HI column density is directly
related to the depth of the well (and to the associated mass).

The trend of increasing correlation with decreasing redshift is also a
strong prediction of any model of structure formation based on
gravitational instability.  On the contrary, theories of explosive
structure formation \cite{vish85,wein89} expect a velocity correlation
either unchanging or diminishing with time.

To be more quantitative an estimate of the masses of the
clouds is needed.
Known the ionization field from the
proximity effect \cite{gial96} and the sizes of the clouds from quasar
pairs and gravitational lenses ($\sim 100 h^{-1}$ Kpc)
\cite{smette92,smette95,bechto94,dinshaw95,dodo2:97}, 
the observed column densities can be transformed in masses
with a photoionization code
\cite{cloudy}.
Then, we can apply a recipe connecting a given model of Universe, the
spectrum of the initial fluctuations, to the clustering of the mass
\cite{matarr97} and, finally, convolve it with the effects of the
non-negligible size of the absorbing structures and their peculiar
velocities $\sigma_v$ \cite{HHW89}.  The results for
two standard cosmological models, a CDM normalized to the abundance of
clusters and a tilted CDM, are shown in Fig. 1 for $\sigma_v \sim 50$ km/s.

\section{Large-scale Clustering}
\subsection{Voids}
A typical way of looking for non-random fluctuations of the line
density on large scales is the search for voids.  Voids in the
Lyman-$\alpha$ forest provide a test for models of the large-scale
structure in the Universe and of the homogeneity of the UV ionizing
flux. Previous searches for megaparsec-sized voids have produced a few
claims \cite{Crotts89,DB91}, but uncertainties in the line statistics
strongly influence the probability estimate \cite{Ostriker88}.  High
resolution data, less affected by blending effects, are ideal also for
the study of voids.  
Detections of voids in individual cases are interesting
but a more general approach, assessing more quantitatively how common
is the phenomenon
and avoiding the pitfalls of ``a posteriori
statistics'', is preferable.  We have adopted the following
procedure: for each of the objects in the extended sample and for
lines with $\log (N_{HI}) \geq 13.3$ we have estimated through
Montecarlo simulations the void size $\Delta r$ (in comoving
coordinates) for which the probability to find at least one void $ \ge
\Delta r$ is $0.05$.  Then we have searched each spectrum for voids of
dimensions equal or larger than $\Delta r$. In 3 out of the 9 cases
for which the density of lines and the absence of spectral gaps allow
a meaningful analysis, at least one void was detected (two voids for
$0055-26$). The binomial probability corresponding to such occurrences
is $8 \cdot 10^{-3}$.

From the spectra it is apparent that the regions corresponding to the
voids are not completely devoid of lines: weak absorptions are
observed within the voids. This agrees with low-redshift observations
\cite{shull96} that have shown that in the local Universe voids are
not entirely devoid of matter.

Even if underdense regions are statistically significant in our
sample, the filling factor is rather low: the voids cover only about
$2$\% of the available line-of-sight path-length, confirming previous
results \cite{CR87}.

\subsection{Over- and under-densities of lines}
Voids are just an extreme case of spectral regions showing an
under-density of lines. Various theoretical reasons prompt to tackle
the issue of over- and under-densities of lines from a more general
point of view: the typical signature of a ``proximity effect'' due to
a foreground quasar is a lack of weak lines in the Lyman forest,
rather than lack of lines in general \cite{DBVict}; the relative
filling factor of under-densities and over-densities may provide a
constraint on the theories of structure formation.

We have analysed the spectra of the extended sample with a
counts-in-cells technique searching for over- and under-densities of
lines with $\log (N_{HI}) \geq 13.3 $ on scales from 10 to 80 Mpc and
comparing the observed counts with Montecarlo simulations in order to
assess the significance of the deviations.  On smaller scales the shot
noise is too large, on larger scales the ``integral constraint'',
forcing the simulated number of lines to be equal to the observed one,
prevents from the possibility of detecting any deviation.  The
threshold to define significant a deviation (for example in excess) in
a given spectral interval has been set in a way that at a given scale
for a given quasar there is a $0.05$ probability of observing at least
one deviation of this type on the whole spectrum for a locally
Poissonian distribution.

5 QSOs out of 15 show at least one over-density in their spectrum at
10 Mpc scales, corresponding to a binomial probability of $6 \cdot
10^{-4}$ of being drawn from a poissonian distribution of lines.  4
QSOs show at least one over-density of lines at 20 and 30 Mpc scales,
corresponding to a binomial probability of $5 \cdot 10^{-3}$.  At
larger scales the number of significant over-densities decreases.
None is observed at 80 Mpc.

Under-densities appear to be roughly as common as over-densities, once
the lack of sensitivity at lower redshifts and smaller scales, due to
the low density of lines, is taken into account.

The existence of a roughly equal number of over and under-densities on
scales $10 - 40$ Mpc is easily understandable in terms of the linear
theory of the evolution of the perturbations, that is a plausible
approximation at such relatively high redshifts: gravity has not yet
had time to give a significant skewness to the (under)over-density
distribution. Besides, almost any hierarchical clustering scenario
would expect that at $z \sim 2 - 4$ gravity has not yet had enough
time to transfer power on 80 Mpc scales and give origin to significant
over or under-densities.

\vfill
\end{document}